\documentclass[12pt]{article}
\usepackage{amsmath}

\newtheorem{assumption}{Assumption}
\newtheorem{theorem}{Theorem}

\usepackage{graphicx,psfrag,epsf}
\usepackage{enumerate}
\usepackage{natbib}
\usepackage{multirow}
\usepackage[plain,noend]{algorithm2e}
\usepackage{bm}
\usepackage{url} % not crucial - just used below for the URL 
\newcommand{\blind}{1}

\usepackage{DNstyle}

\begin{document}

\def\spacingset#1{\renewcommand{\baselinestretch}%
	{#1}\small\normalsize} \spacingset{1}

%%%%%%%%%%%%%%%%%%%%%%%%%%%%%%%%%%%%%%%%%%%%%%%%%%%%%%%%%%%%%%%%%%%%%%%%%%%%%%

\if1\blind
{
	\title{\bf  Analysis of ``Learn-As-You-Go'' (LAGO) Studies}
	\author{Daniel Nevo\thanks{
			This work was supported by the Director's Office and the National Institute of Environmental Health Sciences,  National Institutes of Health (DP1ES025459) and by the National Institute of Allergy and Infectious Diseases, National Institutes of Health  (R01AI112339).  The BetterBirth study  was supported with funding from the Bill \& Melinda Gates Foundation. The article contents are the sole responsibility of the authors and may not necessarily represent the official views of the Bill \& Melinda Gates Foundation. The authors thank Dr. Katherine Semrau for her assistance with sharing and interpreting the results from the BetterBirth Study.}\hspace{.2cm}\\
	 Department of Statistics and Operations Research, \\Tel Aviv University, Tel Aviv 69978, Israel\\
	and Harvard T.H. Chan School of Public Health, USA\\
		Judith J Lok \\
	 Department of Mathematics and Statistics, \\ Boston University, Boston, MA 02215, USA\\
	 		and Harvard T.H. Chan School of Public Health \\
			and \\
		Donna Spiegelman\\
 Department of Biostatistics and\\
		 Center for Methods in Implementation and Prevention Science (CMIPS)­,\\
		    	Yale School of Public Health, New Haven, CT 06520, USA\\
		    			and		Harvard T.H. Chan School of Public Health, USA\\
	}
	\maketitle
} \fi

\if0\blind
{
	\bigskip
	\bigskip
	\bigskip
	\begin{center}
		{\LARGE\bf Analysis of ``Learn-As-You-Go'' (LAGO) Studies}
	\end{center}
	\medskip
} \fi

\bigskip

\begin{abstract}
	In learn-as-you-go (LAGO)  adaptive studies, the intervention is a complex package consisting of multiple components, and is adapted in stages during the study based on past outcome data. This design formalizes standard practice, and desires for practice, in public health intervention studies. An effective intervention package is sought, while minimizing intervention package cost. When analyzing data from a learn-as-you-go study, the interventions in later stages depend upon the outcomes in the previous stages, violating standard statistical theory. We develop methods for estimating the intervention effects in a LAGO study.  We prove consistency and asymptotic normality using a novel coupling argument, ensuring the validity of the test for the hypothesis of no overall intervention effect. We  develop a confidence set for the optimal intervention package and confidence bands for the success probabilities under alternative package compositions. We illustrate our methods in the BetterBirth Study, which aimed to improve maternal and neonatal outcomes among 157,689 births in Uttar Pradesh, India through a complex, multi-component intervention package.
\end{abstract}

\noindent%
\vfill

\newpage
\spacingset{1.45} % DON'T change the spacing!

\section{Introduction}
\label{Sec:intro}
Adaptive designs have been developed and have been available for use in clinical trials for decades.  The U.S. Food and Drug Administration defines an adaptive design as ``...a clinical study design that allows for prospectively planned modifications based on accumulating study data without undermining the study's integrity and validity'' \citep{FDA}.

The existing literature on adaptive designs has thus far considered several types of prospectively planned design modifications, including blinded sample size reassessment, group sequential testing, interim analysis for benefit or futility, successive re-randomization, changing subgroup proportions or eligibility criteria of the trial \citep{rosenblum2011optimizing} and  dropping treatment arms. Prominent among the techniques developed to preserve the validity of statistical inference when design adaption has occurred is the conditional error function \citep{proschan1995designed,muller2001adaptive,muller2004general}, and combination functions have been used to aggregate $p$-values from multiple stages \citep{bauer1994evaluation,brannath2002recursive}.  See  \cite{kairalla2012adaptive,bauer2016twenty} for recent comprehensive reviews of adaptive designs in clinical trials. In addition to valid  testing,  methods have been developed for estimation in an adaptive group sequential design \citep[e.g. ][]{gao2013exact}.

The present work is motivated by large-scale public health intervention studies of complex multi-component intervention packages. In the newly proposed  ``learn-as-you-go'' (LAGO) design, the intervention, which can e.g. be a treatment, a device, a new way to organize care, or, more likely, a combination thereof, is composed of several components. While subject matter experts have some knowledge with regard to the preferred intervention package, in LAGO, optimal development of the intervention package is an inherent part of the study goals. A LAGO study is conducted in stages. After each stage, the data collected so far are analyzed, the intervention package is reassessed, and a revised  intervention package is rolled out in the next stage. Unlike previous adaptive designs, in the LAGO design, the composition of the intervention package in later stages depends on the outcomes from previous stages.

The Sequential Multiple Assignment Randomized Trial (SMART) design \citep{murphy2005experimental,murphy2007developing} randomizes study participants at more than one time point to pre-specified randomization options with probabilities that depend on participant's past characteristics and outcomes.   The aim of a SMART trial is to estimate the optimal sequence of treatments given patient's covariate and response histories up to the present. It is a \textit{non-adaptive} design method which optimizes a personalized and dynamic intervention, in part by restricting randomization options at each step. In contrast, LAGO identifies a complex \textit{static},  possibly `cluster-personalized', intervention package where, unlike in SMART, the options are unknown at the start of the trial and are estimated anew as a result of trial data up to the current stage. In addition, LAGO studies will add new centers, with new participants, entering at each stage, while in SMART the same individuals are repeatedly re-randomized. Both design strategies are well suited for complex multi-component interventions.

The multiphase optimization strategy \citep[MOST,][]{collins2007multiphase,collins2014optimization} consists of three phases: preparation, optimization and evaluation. The optimal intervention package is developed during the optimization phase, followed by its formal statistical evaluation  in a randomized controlled trial. The aim of MOST is similar to LAGO: to develop  an optimal intervention package and estimate its impact. However, in MOST, the outcomes of the past are used at most in one stage, to determine the optimal package in the optimization phase. The resulting package is then independently studied through a controlled trial in the evaluation phase, using no prior data.

At face value, phase I dose-finding studies have perhaps the greatest similarity to the LAGO design paradigm. In dose-finding studies, the goal is to find the maximum tolerated dose, that is, the highest dose of a drug such that adverse effects of the drug are below a pre-determined threshold.  Dose values are assigned to patients in a sequential manner, and in each step a decision is made to stop and declare that the  maximum tolerated dose has been found, or to continue, and if so, with which dose. The more traditionally used methods include the ``3 + 3'' and ``accelerated titration'' designs \citep{simon1997accelerated,wong2016changing}.  Another popular method is the continual reassessment method  \citep{o1990continual,o1996continual}, which assigns  each patient the current estimated maximum tolerated dose. Methods were also  developed for the optimal dose of two drugs simultaneously  \citep{thall2003dose,wang2005two}.  \cite{rosenberger2002competing} provide a review of the continual reassessment method and additional statistical methods for dose finding studies. Dose-finding studies are generally too small for the application of asymptotic statistical methods, and typically Bayesian approaches have been used.
In contrast, in public health intervention studies, the magnitude of the per-stage sample size is typically much larger than the sample size in dose-finding studies, while the maximum number of stages will be limited.
Additionally, unlike dose-finding studies, where methods are considered for a single or at most dual treatments, the complex public health interventions motivating the development of the LAGO design feature multiple components, some of which are continuous, while others are binary \citep{hallberg2015complex}.

An ad hoc example of a precursor to a formal LAGO study is the ``BetterBirth Study'' \citep{hirschhorn2015learning,semrau2017outcomes} of Ariadne Labs, led by Atul Gawande \citep{gawande2014being}, a joint center of  the Brigham and Women's Hospital and the Harvard T.H. Chan School of Public Health. The BetterBirth Study assessed the use of the World Health Organization's (WHO) Safe ChildBirth checklist, a 31-item checklist of best labor and delivery practices believed to be feasible in resource-limited settings, to reduce maternal and neonatal mortality. The intervention was adapted and tested in a three phase process in Uttar Pradesh, India, where neonatal mortality is 32 per 1000 live births and maternal mortality is 258 per 100,000 births \citep{semrau2017outcomes}. During the first two phases,  the intervention was adapted, and a final version was tested in a cluster randomized trial. This work was generously funded through a grant from the Bill \& Melinda Gates Foundation, and included 157,689  mothers and newborns.

The first  goal of a LAGO study is to identify the optimal intervention package such that the cost of the intervention is minimized and the probability of a desired binary outcome  is above a given threshold. For example, in the BetterBirth Study, the outcome  could be the use of the  WHO Safe ChildBirth checklist, with the aim being, for example, that the checklist is used during at least 90\% of the births. In the illustrative example included in this paper, we investigate a process outcome, oxytocin administration after delivery, with the aim being that 85\% of mothers will receive oxytocin after delivery, as recommended by the WHO, as a proven intervention for preventing postpartum hemorrhage. We determine whether the use of  a multiple component  intervention package that includes on-site coaching visits and an intervention launch of a particular duration,  increases the administration of oxytocin, compared to standard of care.

The second goal of a LAGO study is to assess the overall impact of the intervention strategy, as well as that of its individual components. We present methodology to achieve both goals.

In a LAGO study, the data are not an independent sample. Beginning with the second stage, the recommended intervention package is itself a random variable that depends on previous outcomes. In the final analysis, a LAGO study uses the data from all stages. When considering the asymptotic behavior of the estimators, we assume that the sample size in each stage  increases at a similar rate. In addition, we assume that the intervention in each stage converges in probability to a constant  as the number of observations in the previous stage goes to infinity. This would happen, under the usual regularity conditions, if the intervention in each stage is based on a maximum likelihood estimator obtained from the data collected in previous stages. 

LAGO studies can be further characterized  by  a key design feature which determines the strength of the causal inferences that can be made. In an \textit{uncontrolled} LAGO study, there are neither baseline data available to permit a quasi-experimented before-after comparison nor randomized or non-randomized planned variation in the implementation of the intervention package. Thus, unplanned variation, which is widespread in large-scale public health interventions, serves as the basis for making causal contrasts. Causal inference methods will thus be needed to adjust for possible confounding bias \citep{hernan2019causal,spiegelman2018evaluating}. In a \textit{controlled} LAGO study, baseline outcome data are collected before the intervention is implemented, or in additional centers in which no intervention was implemented. These additional centers may be randomized or not,  to be included in the study as controls. In the design where baseline data serves as the control, the quasi-experimental before-after design serves as the basis for causal contrasts. If, instead or in addition, there are concurrent control centers, stronger causal inference is permitted by design, with the strongest design being a randomized controlled before-after set-up. 

We propose estimators for a LAGO study allowing for several stages, multiple centers or sites,  multiple component complex interventions,  and center-specific baseline covariates that affect the outcome rate, or random center-specific deviations from the recommended intervention, or both. We  show that even in this setup, the optimal intervention can be learned from the combined data from all stages. Even when the optimal intervention in the last stage does not achieve the pre-specified study goal, the optimal intervention is estimated. We  prove consistency and asymptotic normality of the new estimators utilizing  a novel coupling argument.  We further establish the validity of tests for an overall intervention effect. In addition, we  develop a confidence set  for the optimal intervention package and confidence bands for the target outcome probability under various observed or hypothesized intervention packages.

\section{LAGO design - theoretical development} 
\label{Sec:lago}
\subsection{Description of the learn-as-you-go design}
\label{SubSec:Descrip}

The methods we develop in this paper cover an arbitrary number of stages, $K$.  At each stage $k$, a version of the intervention package is implemented in each of $J_k$ centers.  Let $n_{jk}$ denote the sample size (e.g. the number of births) in the $j$-th center at stage $k$. We assume that each center is included in one stage only.  In a randomized controlled trial, centers may be randomized to either intervention or control. Alternatively, data might be collected pre and post the implementation of the  intervention package and then a center  contributes data to both the intervention and the control. 

Asymptotic theory is developed for the setting where the number of patients per center goes to infinity at the same rate in all stages, leading to reliable approximations when	 the number patients in each center is relatively large.  Let $n_k=\sum_{j=1}^{J_k}n_{jk}$ be the number of participants in stage $k$ and  $n=\sum_{k=1}^{K}n_{k}$ be the total number of participants. Our asymptotic inference assumes that the ratio between the number of patients in each center and the total sample size $n$ converges to a constant, and we write $\alpha_{jk}=\lim\limits_{n\rightarrow\infty}n_{jk}/n$; then, $\sum_{k=1}^{K}\sum_{j=1}^{J_k}\alpha_{jk}=1$. Define also $\bar{n}_k=(n_1,...,n_k)$. Proofs are given in Sections 1 and 2  of the supplementary materials.  For ease of presentation, we first develop methodology for a LAGO study consisting of two stages.  Section 3 of the supplementary materials covers studies with more than two stages.

The multivariate intervention package  consists of $p$ components. Let $\mathcal{X}$ be the support of the intervention, that is, all possible intervention values. For example, if all  $p$ intervention components are continuous and each is constrained to be within  a given interval $[\mathcal{L}_r,\mathcal{U}_r],r=1,...,p$, then $\mathcal{X}=[\mathcal{L}_1,\mathcal{U}_1]\times[\mathcal{L}_2,\mathcal{U}_2]\times\cdot\cdot\cdot\times[\mathcal{L}_p,\mathcal{U}_p]$.  Throughout this paper, as would ordinarily be the case in practice, we assume that $\mathcal{X}$ is bounded. 

 For stage 1, an initial $\bx^{(1)}$ (or $\bx^{(1)}_j$ for each center $j$)  is chosen by the investigators, based on their best judgment.   We distinguish between the \textit{recommended} intervention  and the \textit{actual} intervention.  In large scale public health settings, the actual intervention,  denoted by $\bA_j$, may differ from the recommended intervention, due to local constraints or preferences. We denote $\bz_j$ for center-specific  characteristics  reflecting  baseline heterogeneity between centers with respect to the outcome of interest and we consider them fixed, i.e., they are not part of the intervention package. For each center, $\bz_j$ could be, for example, the district of the health center or its monthly birth volume. 
 
 We assume that the probability of success for a single unit $i$ (e.g., participant or birth)  in a center $j$ with characteristics $\bz_j$ under intervention $\bA=\ba_j$,
 $p_{\ba_j}(\bbeta;\bz)=pr(Y_{ij}=1\mid\bA_j=\ba_j,\bX_j=\bx_j,\bz_j;\bbeta)$, does not depend on the recommended intervention $\bx_j$, except through the actual intervention $\ba_j$, and follows a logistic regression model
 \begin{equation}
 \label{Eq:LogitDefMulti}
 \logit p_{\ba_j}(\bbeta;\bz_j)=  \beta_0+\bbeta^T_1\ba_j + \bbeta^T_2\bz_j,
 \end{equation}
 where $\bbeta^T=(\beta_0, \bbeta^T_1, \bbeta^T_2)$ is a vector of unknown parameters, such that $\bbeta_1$ describes the effects of the $p$ intervention package components. For centers in the control arm or for pre-intervention data, if available, $\ba=\bx=\boldsymbol{0}$.  We assume that in each stage, conditionally on all $\ba_j$ and $\bz_j$,   outcomes are independent within and between centers. Learning the intervention, however, causes dependence between stages, which we consider below.
 
A main goal of the LAGO design is to identify the optimal intervention package. Let $\tilde{p}$ be a pre-specified  outcome probability goal and $C(\bx)$ be a known cost function.  For example, in the BetterBirth Study, one may want to find the minimal number of on-site coaching visits to ensure that oxytocin is administrated to the mother right after delivery in at least 85\% of births ($\tilde{p}=0.85$).  
  If $\bbeta$ were known, an optimal  intervention for a  center with covariates $\bz_j$ could be the solution to the center-specific optimization problem
 \begin{equation}
 \label{Eq:OptimDefCener}
 \min_{\bx_j} C(\bx_j) \quad \text{subject to} \quad p_{\bx_j}(\bbeta;\bz_j)\ge \tilde{p} \quad \& \quad \bx_j \in \mathcal{X}.
 \end{equation}
 Computational issues regarding solving \eqref{Eq:OptimDefCener} will be discussed in Section \ref{SubSec:Compute}. We assume that for the true parameter values, there is a unique solution to \eqref{Eq:OptimDefCener}.  For example, if the intervention has two components with unit costs $c_1$ and $c_2$ and a linear cost function, we assume that $\beta_{11}/c_1 \ne \beta_{12}/c_2$. 
 Alternatively, other optimization criterion can be considered. For example,  the optimal intervention could require that the intervention results in an outcome probability $\tilde{p}$ when calculating a weighed average over a group of centers $\{j=1,...,J\}$, with sample sizes $n_j$. That is,
 $$
 \min\limits_{\bx_1,....,\bx_{J}} \sum_{j=1}^{J}C(\bx_j) \quad \text{subject to} \quad \frac{1}{N}\sum_{j=1}^{J}n_jp_{\bx_j}(\bbeta;\bz_j)\ge \tilde{p} \quad \& \quad \bx_{j} \in \mathcal{X} \quad \forall j
 $$
 where $N=\sum_{j=1}^{J}n_j$. In this paper we focus on  \eqref{Eq:OptimDefCener}.
 
  We continue our description of the data and model.  Let $\bar{\bz}^{(k)}=(\bz^{(k)}_1,...,\bz^{(k)}_{J_1})$ be the  observed center characteristics in each of the $J_k$ stage $k$ centers. We start with stage 1.   Let $\bx^{(1)}_j$  be the recommended (multivariate) intervention package for center $j$ in stage 1, which in the absence of $\bz$, may be the same for all centers. We assume that the stage 1 recommended interventions $\bx^{(1)}_j$, $j=1,..,J_1$, are determined before the trial starts.
The actual intervention in center $j$ of stage 1 is, however, $\ba^{(1)}_{j}=h^{(1)}_{j}(\bx^{(1)}_{j})$, where $h^{(1)}_{j}$ is a deterministic center-specific continuous function  from $\mathcal{X}$ to $\mathcal{X}$   that determines how center $j$ implements the actual intervention based on the recommendation $\bx^{(1)}_{j}$. We do not require that the $h^{(1)}_{j}$ are known, but only that the $\ba^{(1)}_{j}$ are observed.
Let $Y^{(1)}_{ij}$ be the binary outcome of interest for patient $i$ in center $j$ of stage 1,  each following model \eqref{Eq:LogitDefMulti}, and let the outcome vector in center $j$ of stage 1 be  $\bY^{(1)}_j=(Y^{(1)}_{1j},...,Y^{(1)}_{n_{j1}j})$.   Let $\bar{\ba}^{(1)}=(\ba^{(1)}_1,...,\ba^{(1)}_{J_1})$ and $\bar{\bY}^{(1)}=(\bY^{(1)}_{1},...,\bY^{(1)}_{J_1})$ be the stage 1 actual interventions and outcomes, respectively.

Following the stage 1 data collection, a stage 1 analysis is conducted to determine the recommended interventions for the  new centers in stage 2, denoted by $\hat{\bx}_{j}^{opt,(2,n_1)}, j=1,...,J_2$. If there are control centers, their recommended intervention and their actual intervention are zero.   The value $\hat{\bx}_{j}^{opt,(2,n_1)}$ is chosen through a function, $g$, that takes as input the stage 1 data, the goal of the intervention, and the center-specific covariates and returns a recommended intervention, which is usually the estimated optimal intervention $\hat{\bx}_{j}^{opt,(2,n_1)}=g(\bar{\ba}^{(1)},\bar{\bY}^{(1)}, \bar{\bz}^{(1)}, \bz^{(2)}_{j})$.  Then,  $\hat{\bx}_{j}^{opt,(2,n_1)}$  can be obtained by solving  the optimization problem given in  \eqref{Eq:OptimDefCener} for each center, with $\bbeta$ replaced by an estimator $\hat{\bbeta}^{(1)}$ based on the  stage 1 data alone. The  superscript, $n_1$, in $\hat{\bx}_{j}^{opt,(2,n_1)}$ reminds us that 
 $\hat{\bx}_{j}^{opt,(2,n_1)}$ is a random variable that is a function of the data from the  $n_1$ participants in stage 1.

The actual intervention implemented in center $j$ of stage 2 is $\bA^{(2,n_1)}_{j}=h^{(2)}_{j}(\hat{\bx}_{j}^{opt,(2,n_1)})$, where $h^{(2)}_{j}$ are the analogues of $h^{(1)}_{j}$, but now for the stage 2 centers.  Let $\bar{\hat{\bx}}^{opt,(2,n_1)}=(\hat{\bx}^{opt,(2,n_1)}_1,...,\hat{\bx}^{opt,(2,n_1)}_{J_2})$ be the recommended interventions at the $J_2$ stage 2 centers. Once  $\bar{\hat{\bx}}^{opt,(2,n_1)}$ are determined,   stage 2 outcomes are collected under the actual interventions  $\bar{\bA}^{(2,n_1)}=(\bA^{(2,n_1)}_1,...,\bA^{(2,n_1)}_{J_2})$, which may be the same as $\bar{\hat{\bx}}^{opt,(2,n_1)}$. Let $\bY^{(2,n_1)}_{j}=(Y^{(2,n_1)}_{1j},...,Y^{(2,n_1)}_{n_{j2}j})$ be the  stage 2 outcomes in center $j$, each following model \eqref{Eq:LogitDefMulti}, and   $\bar{\bY}^{(2,n_1)}=(\bY^{(2,n_1)}_{1},...,\bY^{(2,n_1)}_{J_2})$ be all the stage 2 outcomes. Our two main assumptions are
\begin{assumption}
	\label{Assump:CondIndep}
	Conditionally on $\bar{\hat{\bx}}^{opt,(2,n_1)}$, $(\bar{\bA}^{(2,n_1)}, \bar{\bY}^{(2,n_1)})$ are independent of the stage 1 data $(\bar{\ba}^{(1)},\bar{\bY}^{(1)})$.
\end{assumption}
\begin{assumption}
	\label{Assump:gFunction}
	For each $j=1,...,J_2$, the  stage 2 recommended intervention 
	$\hat{\bx}_{j}^{opt,(2,n_1)}$ converges in probability to a center-specific limit $\bx^{(2)}_{j}$.
\end{assumption}
Assumption \ref{Assump:CondIndep} assumes that learning takes place only through the determination of the recommended intervention.
It  ensures that the dependence between the stage 1 data and stage 2 outcomes is solely due to the dependence of the $\hat{\bx}_{j}^{opt,(2,n_1)}$ on the stage 1 data. It specifically means that, given $\bar{\hat{\bx}}^{opt,(2,n_1)}$, the actual intervention in a stage 2 center is conditionally independent of $\bar{\bY}^{(1)}$. Under  Assumption 	\ref{Assump:CondIndep}, and the aforementioned assumption that conditionally on the actual interventions, the outcomes do not depend on the recommended interventions,  we can conclude that  in stage 2,  $pr(\bar{\bY}^{(2,n_1)}\mid \bar{\bA}^{(2,n_1)},\bar{\hat{\bx}}^{opt,(2,n_1)},\bar{\bz}^{(2)},\bar{\bY}^{(1)})=
pr(\bar{\bY}^{(2,n_1)}\mid  \bar{\bA}^{(2,n_1)},\bar{\bz}^{(2)})$, so the logistic regression model \eqref{Eq:LogitDefMulti} holds for the stage 2 data.
Assumption	\ref{Assump:gFunction} implies that in the presence of more and more stage 1 data under $\ba_{j}^{(1)}, j=1,...,J_1$, each of the estimated optimal intervention packages $\hat{\bx}_{j}^{opt,(2,n_1)}, j=1,...J_2$, converges in probability to a fixed value $\bx^{(2)}_{j}$. For example, Assumption \ref{Assump:gFunction} will hold if $\bar{\hat{\bx}}^{opt,(2,n_1)}$ are continuous functions of the stage 1 maximum likelihood estimator, $\hat{\bbeta}_1$, as is the case if $\hat{\bx}_{j}^{opt,(2,n_1)}$ solves \eqref{Eq:OptimDefCener} and  $\beta_{11}/c_1 \ne \beta_{12}/c_2$.   Under Assumption	\ref{Assump:gFunction}   and continuity of the $h_j$'s, the Continuous Mapping Theorem implies that $\bA_{j}^{(2,n_1)}=h^{(2)}_{j}(\hat{\bx}_{j}^{opt,(2,n_1)})$ converges in probability to $\ba^{(2)}_{j}=h^{(2)}_{j}(\bx^{(2)}_{j})$.  

In fact, the results we prove in this paper regarding the estimators obtained at the end of the study hold not only for $g(\bar{\ba}^{(1)},\bar{\bY}^{(1)}, \bar{\bz}^{(1)}, \bz^{(2)}_{j})=\hat{\bx}_{j}^{opt,(2,n_1)}$, but  under any choice of function $g$ for the recommended  intervention, as long as Assumption \ref{Assump:gFunction} holds. Further details about $g$ and proofs of this claim are given in Section 2 of the supplementary materials. 

\subsection{$\hat{\bbeta}$ and its asymptotic properties}
\label{SubSec:MLE}
We estimate $\bbeta$ after the $K$ stages are concluded. As in previous sections,  for ease of development,  we consider here $K=2$.  Section 3 of the supplementary materials covers the case of $K>2$. 

We propose to estimate $\bbeta$ by solving the estimating equations
\begin{align}
\begin{split}
	\label{Eq:UbetaMulti}
	0=\bU(\bbeta)&=\frac{1}{n}
	\left\{\sum_{j=1}^{J_1}\sum_{i=1}^{n_{j1}}\begin{pmatrix} 1\\ \ba\textbf{}^{(1)}_j\\ \bz^{(1)}_j \end{pmatrix}\bigg(Y^{(1)}_{ij}-p_{\ba^{(1)}_{j}}(\bbeta;\bz^{(1)}_j)\bigg)\right.\\
	&\left.+\sum_{j=1}^{J_2}\sum_{i=1}^{n_{j2}}\begin{pmatrix} 1\\ \bA^{(2,n_1)}_{j}\\ \bz^{(2)}_{j} \end{pmatrix}\bigg(Y^{(2,n_1)}_{ij}-p_{\bA^{(2,n_1)}_{j}}(\bbeta;\bz^{(2)}_{j})\bigg)\right\}.
\end{split}
\end{align}
In Section 2.1 of the  supplementary materials, we show that the estimator $\hat{\bbeta}$ that solves \eqref{Eq:UbetaMulti} is also a maximum partial likelihood estimator, although it is not needed for the proofs below. These estimating equations \eqref{Eq:UbetaMulti} also arise if the interventions were determined a priori, so $\hat{\bbeta}$ can be estimated using standard software.

Asymptotic theory  for $\hat{\bbeta}$  is complicated, however, by the fact that $\bar{\bY}^{(1)}$ and $(\bar{\bA}^{(2,n_1)}, \bar{\bY}^{(2,n_1)})$ are not independent. Thus, the score function, $U(\bbeta)$, is not a sum of independent random variables.

 Let $\mathcal{B}$ be the parameter space for $\bbeta$. A conditional expectations argument (Equation (A.9) in the supplementary materials) shows  that the score function has mean zero when evaluated at the true value, denoted by $\bbeta^\star$. Furthermore, we show in the supplementary materials (Equation (A.10)) that the two terms in \eqref{Eq:UbetaMulti}, although dependent, are uncorrelated. 
These two properties are useful for proving that  $\hat{\bbeta}$ is consistent: 
\begin{theorem}
	\label{them:consis}
		(Consistency)
	Assume $\mathcal{B}$ is compact. 	Under Assumptions \ref{Assump:CondIndep} and \ref{Assump:gFunction},  $\hat{\bbeta}\xrightarrow{P} \bbeta^\star$.
\end{theorem}
The proof is given in Section 2.2 of the supplementary materials.

Asymptotic normality also poses a challenge due to the dependence between the two summands in  $\bU(\bbeta)$. It can be shown that $\partial\bU(\bbeta)/\partial{\bbeta}$ converges in probability to $-I(\bbeta)$, for all $\bbeta\in\mathcal{B}$, with $I(\bbeta)$ given  in Section 2.3 of the supplementary materials. The following theorem establishes asymptotic normality of $\hat{\bbeta}$:
\begin{theorem}
	(Asymptotic normality) Under Assumptions  \ref{Assump:CondIndep} and \ref{Assump:gFunction},
	\begin{equation}
		\label{Eq:AsyNorm}
		n^{1/2}(\hat{\bbeta}-\bbeta^\star)\xrightarrow{\mathcal{D}} N\left(0, I^{-1}(\bbeta^\star)\right).
	\end{equation}
\end{theorem}
The full proof of Theorem 2 is given in Section 2.3 of the supplementary materials. Here we outline the main parts of the proof, which rests upon a novel coupling argument. First, by the mean value theorem  and further arguments, it can be shown that  the asymptotic distribution of $n^{1/2}(\hat{\bbeta}-\bbeta^\star)$ is the same as the asymptotic distribution of 
\begin{align}
	\begin{split}
		\label{Eq:betahatsame}
		\left[I(\bbeta^\star)\right]^{-1}&n^{-1/2}\left[\sum_{j=1}^{J_1}\sum_{i=1}^{n_{j1}}\begin{pmatrix} 1\\ \ba^{(1)}_j\\ \bz^{(1)}_j \end{pmatrix}\bigg(Y^{(1)}_{ij}-p_{\ba^{(1)}_{j}}(\bbeta^\star;\bz^{(1)}_{j})\bigg)\right.\\
		&+\left.\sum_{j=1}^{J_2}\sum_{i=1}^{n_{j2}}\begin{pmatrix} 1\\ \bA^{(2,n_1)}_{j}\\ \bz^{(2)}_{j} \end{pmatrix}\bigg(Y^{(2,n_1)}_{ij}-p_{\bA^{(2,n_1)}_{j}}(\bbeta^\star;\bz^{(2)}_{j})\bigg)\right].
	\end{split}
\end{align}
We next show that the asymptotic distribution of the part of \eqref{Eq:betahatsame} that does not involve $I(\bbeta^\star)$ is multivariate normal.  The following coupling argument deals with the fact that the two summands in 		\eqref{Eq:betahatsame} are not independent.  For each $j=1,...,J_2$, let $Y^{(2)}_{ij}$, $i=1,...,n_{j2}$, be independent Bernoulli random variables,  independent of all stage 1 data, with success probability $p_{\ba^{(2)}_{j}}(\bbeta^\star;\bz^{(2)}_{j})$, where, as defined before, $\ba^{(2)}_{j}=h^{(2)}_{j}(\bx^{(2)}_{j})$. We construct variables $\tilde{Y}^{(2,n_1)}_{ij}$ which, given the stage 1 data and the $\bA_{j}^{(2,n_1)}$, have the same distribution as the original $Y^{(2,n_1)}_{ij}$, but coupled (see e.g. \cite{lindvall2002lectures}) with  the $Y^{(2)}_{ij}$ in the following way.
Let $W_{ij}$ be independent uniform $(0,1)$ random variables, independent of all other variables introduced so far. For the case $p_{\ba^{(2)}_{j}}(\bbeta^\star;\bz^{(2)}_{j})>p_{\bA^{(2,n_1)}_{j}}(\bbeta^\star;\bz^{(2)}_{j})$, let
$$
\tilde{Y}^{(2,n_1)}_{ij} = \left\{\begin{array}{cl} 0 & {\rm if}\; Y_{ij}^{(2)}=0\\
0 &  {\rm if}\; Y_{ij}^{(2)}=1 \;{\rm and}\;W_{ij}< \frac{p_{\ba_{j}^{(2)}}(\bbeta^\star;\bz^{(2)}_{j})-p_{\bA_{j}^{(2,n_1)}}(\bbeta^\star;\bz^{(2)}_{j})}{p_{\ba_{j}^{(2)}}(\bbeta^\star;\bz^{(2)}_{j})}\\
1 &  {\rm if}\; Y_{ij}^{(2)}=1 \;{\rm and}\;W_{ij}\geq \frac{p_{\ba_{j}^{(2)}}(\bbeta^\star;\bz^{(2)}_{j})-p_{\bA_{j}^{(2,n_1)}}(\bbeta^\star;\bz^{(2)}_{j})}{p_{\ba_{j}^{(2)}}(\bbeta^\star;\bz^{(2)}_{j})}.\\
\end{array}\right.
$$
A similar expression is given in the supplementary materials for the case  $p_{\ba_{j}^{(2)}}(\bbeta^\star;\bz^{(2)}_{j})\leq p_{\bA_{j}^{(2,n_1)}}(\bbeta^\star;\bz^{(2)}_{j})$. 
The key property of the coupling argument is that given $\bA^{(2,n_1)}_{j}$ and the stage 1 data, the distribution of the coupled $\tilde{Y}^{(2,n_1)}_{ij}$  is identical to the distribution of the original $Y^{(2,n_1)}_{ij}$. Therefore, when we replace $Y^{(2,n_1)}_{ij}$ with $\tilde{Y}^{(2,n_1)}_{ij}$ in \eqref{Eq:betahatsame}, the distribution of \eqref{Eq:betahatsame} is unaffected. In the supplementary materials, we use the coupled outcomes to show that  the part of \eqref{Eq:betahatsame} that does not involve $I(\bbeta^\star)$ has the same asymptotic distribution as
\begin{equation}
\label{Eq:FinalExpress}
 n^{-1/2}\Bigg\{\sum_{j=1}^{J_1}\sum_{i=1}^{n_{j1}}\begin{pmatrix} 1\\ \ba^{(1)}_j\\ \bz^{(1)}_j \end{pmatrix}\big(Y^{(1)}_{ij}-p_{\ba^{(1)}_{j}}(\bbeta^\star;\bz^{(2)}_{j})\big) +	 \sum_{j=1}^{J_2}\sum_{i=1}^{n_{j2}}\begin{pmatrix} 1\\ \ba^{(2)}_{j}\\ \bz^{(2)}_{j} \end{pmatrix}\big(Y^{(2)}_{ij}-p_{\ba^{(2)}_{j}}(\bbeta^\star;\bz^{(2)}_{j})\big)\Bigg\}.
\end{equation}
The outcomes $\bar{\bY}^{(1)}$ and $\bar{\bY}^{(2)}=(\bar{\bY}^{(2)}_{1},...,\bar{\bY}^{(2)}_{J_2})$ are independent, because the $\bY^{(2)}_{ij}$ are the outcomes under the constant intervention $\ba^{(2)}_{j}$. Therefore, by standard logistic regression theory, the expression in \eqref{Eq:FinalExpress} converges in distribution to a normal random variable with mean zero and variance $I(\beta^\star)$.
Combining the asymptotic normality of \eqref{Eq:FinalExpress} with \eqref{Eq:betahatsame} implies that Theorem 2 holds.

The asymptotic variance can be consistently estimated from the data by  replacing $\ba^{(2)}_{j}$, $\bbeta^\star$, $\alpha_{j1}$ and $\alpha_{j2}$  with $\bA^{(2,n_1)}_{j}$,  $\hat{\bbeta}$, $n_{j1}/n$ and $n_{j2}/n$, respectively, in $I(\bbeta^\star)$. The asymptotic variance and its approximation are the same as if the  interventions  were fixed in advance and $\bar{\bY}^{(1)}$ and $\bar{\bY}^{(2,n_1)}$ were independent.

\subsection{Hypothesis testing}
\label{SubSec:Testing}
A major goal of a LAGO study is to test the null hypothesis of no overall intervention effect. One way to test this is to carry out a  test for the subvector of $\bbeta$ characterizing the effect of the intervention. That is,  to test  $H_0:\bbeta_1=0$ in model \eqref{Eq:LogitDefMulti} using the asymptotic normality result of Section \ref{SubSec:MLE}. Because of this asymptotic normality result, the Wald  or  likelihood ratio tests for $H_0:\bbeta_1=\bbeta_1^0$ are asymptotically valid for any constant $\bbeta_1^0$. 
 
 Alternatively, in a controlled LAGO design, let $Q$ be a group indicator that equals one for the intervention group and zero for the control, and let $p_0$  and $p_1$ be the  success probabilities under $Q=0$ and $Q=1$, respectively. Then, an alternative test for an overall intervention  effect,   $H_0:\bbeta_1=0$, can be carried out by testing $H_0: p_0=p_1$. The latter test is valid  despite the   adaption of the intervention package.   By Assumption \ref{Assump:CondIndep}, the  dependence between the stage 2 and stage 1 data is solely due to the stage 1 data determining  the stage 2 recommended intervention, which, in turn, affects the actual stage 2 intervention,  and thus  the  stage 2 outcomes. However, under the null, there is no effect of the actual intervention on the stage 2 outcomes.  Therefore, under the null, regardless of the way the intervention was adapted, the stage 1 and stage 2 outcomes are independent. Thus, a standard test for equal probabilities  in the control and the intervention arms is valid. While not needed due to our asymptotic results, the same arguments could have been used for the standard tests of  $H_0:\bbeta_1=0$. 
  
In a controlled LAGO design, an alternative, possibly more powerful, test for the overall effect of the intervention in the presence of center  characteristics  is to consider $H_0:\gamma=0$ in the model $\logit \tilde{p}_{Q}(\bbeta,\gamma;\bz)=  \bbeta_0+\bbeta^T_2\bz + \gamma Q$. As before, in light of the between-stages independence under the null,   $\bbeta_1=0$ in model \eqref{Eq:LogitDefMulti} implies $\gamma=0$. 

\subsection{Confidence sets and confidence bands}
\label{SubSec:Conf}
After the conclusion of the study, the  optimal intervention is estimated as the solution to \eqref{Eq:OptimDefCener} with $\bbeta$ replaced by $\hat{\bbeta}$. To obtain an asymptotic 95\% confidence set for the optimal intervention $\bx^{opt}$, we first  obtain a confidence interval for $p_{\bx}(\bbeta^\star;\tilde{\bz})$, for a given  $\bz=\tilde{\bz}$ and for each $\bx\in \mathcal{X}$. To do this, we calculate  a 95\% confidence interval for $\logit(p_{\bx}(\bbeta^\star;\tilde{\bz}))$, i.e., for $(1 \; \bx^T \; \tilde{\bz}^T){\bbeta^\star}$:
$$
CI_{\bx}= \left[(1 \; \bx^T \; \tilde{\bz}^T)\hat{\bbeta}-1.96\sigma(\hat{\bbeta};\bx,\tilde{\bz}), \qquad (1 \; \bx^T \; \tilde{\bz}^T)\hat{\bbeta} +1.96\sigma(\hat{\bbeta};\bx,\tilde{\bz})\right],
$$
where $\sigma^2(\hat{\bbeta};\bx,\tilde{\bz})= (1 \; \bx^T \; \tilde{\bz}^T)n^{-1}\hat{I}^{-1}(\hat{\bbeta})(1 \; \bx^T \; \tilde{\bz}^T)^T$  is the estimated variance of $(1 \; \bx^T \; \tilde{\bz}^T)\hat{\bbeta}$, and $n^{-1}\hat{I}^{-1}(\hat{\bbeta})$ is the estimated variance of $\hat{\bbeta}$.
The 95\% confidence interval for $p_{\bx}(\bbeta^\star;\tilde{\bz})$ is $CI_{p_{\bx}}=\expit(CI_{\bx})$. Then, we obtain the confidence set for the optimal intervention as $CS(\bx^{opt})=\{\bx:CI_{p_{\bx}} \ni \tilde{p}\}$. That is, $CS(\bx^{opt})$ includes  intervention packages for which $\tilde{p}$ is inside the confidence interval for the success probability under those interventions.

We now show that the confidence set $CS(\bx^{opt})$ contains $\bx^{opt}$ with the specified probability of 0.95. Recall that under the assumption that $\tilde{p}$ can be achieved, $p_{\bx^{opt}}(\bbeta^\star;\tilde{\bz})=\expit[{(1 \; {\bx^{opt}}^T \; \tilde{\bz}^T)\bbeta^\star}]=\tilde{p}$. Therefore,
$$
pr(CS(\bx^{opt}) \ni \bx^{opt})=Pr(CI_{p_{\bx^{opt}}} \ni \tilde{p})=Pr(CI_{p_{\bx^{opt}}} \ni p_{\bx^{opt}}(\bbeta^\star;\tilde{\bz}))=0.95.
$$
Implementing this procedure is simple and its calculation is fast. Because  calculating $CS(\bx^{opt})$ does not depend upon estimating $\bx^{opt}$, it does not involve the optimization algorithm.

At the end of the study,  researchers might be interested in a variety of potential intervention packages in $\mathcal{X}$  that were not necessarily identified as of interest a priori. We propose a method to develop confidence bands for the outcome probabilities $p_{\bx}(\bbeta;\tilde{\bz})$ for a range of $\bx\in\mathcal{X}$ of interest, simultaneously. These confidence bands allow researchers to study the entire intervention space when comparing potential choices of the intervention package. We propose  a procedure that is based on the asymptotic normality of $\hat{\bbeta}$ and on Scheff\'e's method \citep{scheffe1959analysis}.
 First, for all $\bx\in \mathcal{X}$, construct $CB_{\bx}$ to obtain 95\% confidence bands for  $\{(1 \; \bx^T \; \tilde{\bz}^T){\bbeta^\star}: \bx\in \mathcal{X}\}$,
\begin{equation*}
CB_{\bx}= \left[(1 \; \bx^T \; \tilde{\bz}^T)\hat{\bbeta}-\chi^2_{0.95,p+q+1}\sigma(\hat{\bbeta};\bx,\tilde{\bz}), \qquad (1 \; \bx^T \; \tilde{\bz}^T)\hat{\bbeta}+\chi^2_{0.95,p+q+1}\sigma(\hat{\bbeta};\bx,\tilde{\bz}),\right],
\end{equation*}
with $\sigma(\hat{\bbeta};\bx,\tilde{\bz})$ defined as before and $\chi^2_{0.95,p+q+1}$  the 95\% quantile of a $\chi^2_{p+q+1}$ distribution.  As before, we transform $CB_{\bx}$ into confidence bands for $p_{\bx}(\bbeta;\tilde{\bz})$ by setting $CB_{p_{\bx}}=\expit(CB_{\bx})$. These confidence bands guarantee asymptotic simultaneous 95\% coverage for all possible intervention package compositions; the proof is given in Section 4 of the supplementary materials. 

\subsection{Computation of the optimal intervention}
\label{SubSec:Compute}
	The algorithm used to solve \eqref{Eq:OptimDefCener} after stage $k$,  using $\hat{\bbeta}^{(k)}$, depends on the form of $C(\bx)$.  Under a linear cost function with unit costs $c_r$ for the $r$--th component of the intervention, the solution is achieved by 1. setting all components to their minimal value $\mathcal{L}_r$, 2. ordering  the components by their estimated cost-efficiency $\hat{\beta}_{1r}/c_r$, and 3. increasing the most cost-efficient  component until either $\tilde{p}$ is achieved  or until this component reaches its maximal value, and then moving to the  next most cost-efficient component among the remaining components. For non-linear cost functions, standard non-linear optimization algorithms can be used.  

\section{Simulations}
\label{Sec:Sims}

We conducted simulation studies to investigate the finite sample properties of our methods.  We simulated 1000 data sets per simulation scenario. We considered a two-stage controlled LAGO design with equal number of centers per stage $J$, with half the centers in the intervention arm and half in the control arm.  The total sample size available at the end of the study is $J(n_{1j} + n_{2j})$.  We considered the values  $J=6,10, 20$, $n_{1j}=50,100,200$, and $n_{2j}=100, 200,500, 1000$. The intervention had two components, $\bx=(x_1, x_2)$, with unit costs $c_1=1$ and $c_2=8$. The minimum and maximum values of $X_1$ and $X_2$ were $[\mathcal{L}_1,\mathcal{U}_1]=[0,2]$  and  $[\mathcal{L}_2,\mathcal{U}_2]=[0,5]$.  We considered the following  values for $\exp(\bbeta^\star_{1})=(\exp(\beta^\star_{11}),\exp(\beta^\star_{12}))$: $(1, 1)$ (the null), $(1, 1.2)$, $(1, 1.5)$, $(1.2, 1.5)$, and $(1.2, 2)$.  A single center covariate $z$ was normally distributed with mean 0 and variance 1 and its coefficient was taken to be $\beta^\star_{2}=\log(0.75)$. For simplicity, we did not include an intercept in model \eqref{Eq:LogitDefMulti}, although each center had its own baseline success probability due to $z$. For $z=0$, the probability of success in the control arm was 0.5.   The stage 2 recommended intervention was based on solving the optimization problem \eqref{Eq:OptimDefCener} using the  stage 1 estimates of $\bbeta$.  Section 5.1 of the supplementary materials provides the details on what was done when no solution existed for which $\tilde{p}$ was reached. 

Selected results are presented in Tables \ref{Tab:SimsBetas} and \ref{Tab:SimsInter}. Table \ref{Tab:SimsBetas} presents results on the performance of $\hat{\bbeta}$, and shows that for $J>6$, the finite sample bias was minimal,  the mean estimated standard error was very close to the empirical standard deviation, and the empirical coverage rate of the confidence intervals for the effects of the individual package components was very close to 95\%.  Moreover, Section 5.2 of the supplementary shows that the type I error rate of the tests discussed in Section \ref{SubSec:Testing} was close  to 0.05.

\begin{table}
	\caption{Simulation study: results for individual  package component effects. Unit costs were $c_1=1$ and $c_2=8$. 
	}\small{\begin{center}%
		\begin{tabular}{lccccccccc}
		%	\\
			$\exp(\bbeta^\star)$ & $n_{1j}$ & $n_{2j}$ & $J$ &  \multicolumn{3}{c}{$\hat{\beta}_{11}$} & \multicolumn{3}{c}{$\hat{\beta}_{12}$} \\
			&&&& \%RelBias & SE/EMP.SD & CP95  &  \%RelBias & SE/EMP.SD & CP95  \\
			&&&&& $(\times100)$ &  & & $(\times100)$ & \\[5pt]
			$(1.2, 1.5)$ & 50 & 100 &  6 & -1.1 & 92.0 & 95.2 & -2.1 & 83.3 & 94.2 \\ 
			&  &  &  10 & -3.0 & 100.1 & 95.6 & -0.8 & 93.4 & 94.9 \\ 
			&  &  & 20 & 0.1 & 103.5 & 95.5 & -0.6 & 104.9 & 96.1 \\ 
			&  & 200 &  6 &   -3.0 & 88.4 & 94.9 & -3.1 & 83.5 & 95.2 \\ 
			&  &  &  10 & -6.6 & 92.9 & 94.5 & -0.9 & 93.5 & 94.9 \\ 
			&  &  & 20 & 0.2 & 102.5 & 95.6 & -0.6 & 97.7 & 95.3 \\ 
			& 100 & 100 &  6 & -0.8 & 89.5 & 95.1 & -1.6 & 86.7 & 95.2 \\ 
			&  &  &  10 & 3.5 & 102.2 & 95.7 & -1.3 & 102.2 & 95.0 \\ 
			&  &  & 20 & 1.5 & 100.7 & 95.3 & -0.4 & 101.1 & 95.2 \\ 
			&  & 200 &  6 & -2.2 & 90.4 & 94.6 & -1.4 & 89.7 & 96.0 \\ 
			&  &  &  10 & -0.8 & 102.7 & 96.7 & -0.7 & 95.9 & 95.5 \\ 
			&  &  & 20 & -0.3 & 97.4 & 94.7 & -0.4 & 96.7 & 94.1 \\ 
			$(1.2, 2)$ & 50 & 100 &  6 & -11.4 & 89.0 & 94.8 & -0.4 & 82.2 & 96.1 \\ 
			& &  &  10 &  -7.3 & 103.7 & 95.7 & 0.4 & 104.4 & 96.5 \\ 
			& &  & 20 &  -3.1 & 99.0 & 94.7 & -0.1 & 100.8 & 95.0 \\ 
			& & 200 &  6 & -15.8 & 92.6 & 95.0 & 1.4 & 89.7 & 94.9 \\ 
			& &  &  10 & -8.1 & 93.3 & 95.7 & 0.3 & 99.6 & 95.5 \\ 
			& &  & 20 &  -1.8 & 100.1 & 95.3 & -0.5 & 102.5 & 96.6 \\ 
			& 100 & 100 &  6 & -6.0 & 96.2 & 96.3 & 0.0 & 94.0 & 95.2 \\ 
			& &  &  10 & -2.7 & 98.2 & 95.1 & -0.2 & 104.7 & 95.4 \\ 
			& &  & 20 & -2.7 & 100.7 & 95.2 & 0.2 & 102.2 & 95.2 \\ 
			& & 200 &  6 & -8.9 & 95.4 & 95.4 & 0.3 & 83.8 & 96.5 \\ 
			& &  &  10 & -5.0 & 95.6 & 94.6 & 0.0 & 97.3 & 95.3 \\ 
			&  &  & 20 & -3.2 & 98.9 & 94.4 & 0.1 & 104.7 & 95.5 \\ 
	\end{tabular}
\end{center}}
	\label{Tab:SimsBetas}
%	\begin{tabnote}
	\footnotesize	\%RelBias, percent relative bias $100(\hat{\beta}-\beta^\star)/\beta^\star$; SE, mean estimated standard error; EMP.SD, empirical standard deviation; CP95, empirical coverage rate of 95\%  confidence intervals.
	%\end{tabnote}
\end{table}
 
  Table \ref{Tab:SimsInter} presents results for the estimated optimal intervention and success probabilities, for $J=20$ and calculated for a typical center with $z=0$; results for  $J=6,10$ are presented in Section 5.2 of the supplementary materials. The finite sample bias and the root mean squared errors of the final $\hat{\bx}^{opt}$ were generally small and decreased as the number of centers per stage and the sample size increased. The nominal coverage rate of the confidence set for ${\bx}^{opt}$ was approximately 95\%, with the set typically including  between 3 to 14 percent of $\mathcal{X}$, as a measure of precision in the scenarios studied.  We also compared the cost of the estimated optimal intervention to the cost of the true optimal intervention and found it to be almost the same for the scenarios presented in Table \ref{Tab:SimsInter}; see Section 5.2 in the supplementary materials.   Table \ref{Tab:SimsInter} also shows that the empirical coverage rate of the confidence bands for  $p_{\bx}(\bbeta^\star;z=0)$ was very close to 95\%.

\begin{table}
	\caption{Simulation study: results for estimated optimal intervention package and coverage of 95\% confidence bands for success probabilities. Unit costs were $c_1=1$ and $c_2=8$. 
		Results presented for $J=20$  centers per stage.
	}{%
					\footnotesize\begin{tabular}{cccccccccc}
			\\
$\exp(\bbeta^\star)$ & $\bx^{opt}$ & $n_{1j}$ & $n_{2j}$  & Bias($x^{opt}_1$)  & Bias($x^{opt}_2$) & RMSE($\bx^{opt}$) & SetCP95 & SetPerc\% & BandsCP95 \\  
			& & & & $(\times100)$ & $(\times100)$ & $(\times100)$ & & & \\[5pt]
$(1, 2)$ & $(0, 3.2)$ & 50 & 100 & 36.4 & -5.0 & 87.3 & 94.8 & 7.6 & 96.9 \\ 
& &  & 500 & 18.6 & -2.4 & 62.0 & 95.2 & 4.1 & 96.8 \\ 
& & 100 & 100 & 22.6 & -2.8 & 69.0 & 94.5 & 6.1 & 96.7 \\ 
& &  & 500 & 9.8 & -1.3 & 45.3 & 94.5 & 3.7 & 97.5 \\ 
$(1.2, 1.5)$ & $(2, 4.5)$ & 50 & 100 &  -8.4 & 2.4 & 48.9 & 94.4 & 13.3 & 96.8 \\ 
& &  & 500 & -1.9 & 0.9 & 25.0 & 94.9 & 7.7 & 95.9 \\ 
& & 100 & 100 & -4.4 & 1.3 & 38.4 & 94.6 & 12.3 & 95.5 \\ 
& &  & 500 & -0.6 & 2.2 & 18.4 & 94.8 & 7.1 & 95.5 \\ 
$(1.2, 2)$ & $(2, 2.6)$ & 50 & 100 & -31.2 & 4.0 & 81.6 & 94.0 & 14.2 & 95.0 \\ 
& &  & 500 & -15.2 & 3.3 & 57.1 & 94.9 & 8.0 & 94.8 \\ 
& & 100 & 100 & -21.8 & 2.7 & 68.3 & 95.1 & 12.4 & 95.4 \\ 
& &  & 500 &  -9.0 & 2.6 & 44.1 & 94.3 & 7.5 & 95.0 \\ 
\end{tabular}}
\label{Tab:SimsInter}
	\footnotesize RMSE, root of mean squared errors $\{\text{mean}(||\hat{\bx}^{opt}-\bx^{opt}||^2)\}^{1/2}$, mean taken over simulation iterations; SetCP95, empirical coverage percentage of confidence set for optimal intervention; SetPerc\%, mean percent of $\mathcal{X}$ covered by the confidence set; BandsCP95, empirical coverage rate of 95\% confidence bands for $\{p_{\bx}(\bbeta;\bz=0): x \in \mathcal{X}\}$. 
\end{table}

\section{Illustrative example}
\label{Sec:BetterBirth}
The BetterBirth Study consisted of three stages. The first two stages were pilot stages used to develop the intervention package. Stage 3 was a randomized controlled trial. The development of the recommended intervention package was conducted qualitatively, as described in \cite{hirschhorn2015learning}, and the intervention package was adjusted after each pilot stage. The results of the randomized controlled trial were presented and discussed in \cite{semrau2017outcomes}.   The number of centers with data on oxytocin administration in the first, second, and third stages was  2, 4 and 30, respectively. In the first two stages, data in each center were collected before and after the intervention was implemented. In  stage 3, there were 15 centers in the control arm  and 15 centers in the intervention arm. In 5 intervention arm centers, outcome data were also collected before the intervention was implemented.

Here, we focus on the binary outcome of oxytocin administration immediately after delivery, as recommended by the WHO \citep{world2012recommendations} to prevent postpartum hemorrhage, a major cause of maternal mortality. The intervention package components were the  duration of the on-site  intervention launch (in days), the number of coaching visits after the intervention was launched,  leadership engagement (non-standardized initial engagement, standardized initial engagement,  and  standardized initial engagement with follow-up visits) and  data feedback (none; ongoing, paper-based; ongoing, app-based). The four components were adapted in a way that resulted in near multicollinearity. Therefore, for illustration purposes, we  considered the first two components only, launch duration and number of coaching visits.   The launch duration was 3 days in stage 1 and 2 days in   stages 2 and 3. Compared to  stage 1, the intensity of coaching visits  was increased in stage 2, and further increased in  stage 3.  For illustrative purposes,  we truncated the data at 40 coaching visits or less. The baseline center  characteristic we included was the approximate monthly birth volume, given that large facilities might be likely  to follow WHO recommendations about oxytocin administration more closely, regardless of the  intervention package implemented. Other available center characteristics, e.g. number of staff nurses, were highly correlated with the monthly birth volume.

Table \ref{Tab:BBres} provides the estimated effects of the intervention package components after each of the stages, using all available data at that point. The sample size in stage 1 was relatively small, explaining the wide confidence intervals  for the odds ratios.  The final results imply that both package components had an effect. Tests for the overall  effect of the package yielded a  highly significant p-value, regardless of the test we used. 

\begin{table}
	\caption{Package component effect estimates and confidence intervals, calculated after each stage.}\small{%
		\begin{tabular}{cccc}
			%\\
			& Stage 1  & Stages 1-2 & Stages 1-3 \\ 
		&	$n_1=73$ & $(n_1+n_2=1780)$ & $(n_1+n_2+n_3=n=6124)$\\
			& OR (CI-OR) &  OR (CI-OR) & OR (CI-OR)    \\[2pt]
Intercept & 1.07 (0.00, 280.80) & 0.10 (0.07,0.15) & 0.10  (0.09,0.11) \\ 
Coaching  Visits  & 7.95 (1.77,73.95) & 1.11 (0.96,1.28) & 1.08 (1.04,1.12) \\ 
\footnotesize{(per 3 visits)} &&&\\
Launch Duration & 1.41 (0.76,2.64) & 2.65 (1.95,3.77) & 2.79 (2.41,3.23) \\ 
\footnotesize{(days)} &&&\\
Birth Volume & 0.37 (0.00,32.33) & 2.11 (1.93,2.33) & 1.94 (1.84,2.06) \\ 
\footnotesize{(monthly, per 100)} &&&\\
		& \multicolumn{1}{c}{$\hat{\bx}^{opt,(2,n_1)}=(1,5)$} & \multicolumn{1}{c}{$\hat{\bx}^{opt,(3,(n_1,n_2))}=(3,1)$} & \multicolumn{1}{c}{$\hat{\bx}^{opt}=(3,1)$}\\
	\end{tabular}}\\
	\label{Tab:BBres}
	\footnotesize	OR, estimated odds ratio $\exp(\hat{\beta})$; CI-OR, 95\% Confidence interval for the odds ratio. In the estimated optimal interventions, the first component is the launch duration (in days) and the second component is    the number of coaching visits .
\end{table}

After consulting with the study investigators, we assigned  unit costs of \$800 per launch day and \$170 per coaching visit. In practice, implementation costs may also depend on center size and, if so, $C(\bx)$ could be replaced with $C_{\bz}(\bx)$. 

The estimation of the optimal intervention package with linear cost $C(\bx)=c_1x_1+c_2x_2$ was conducted as in the simulation study.  Assuming that at least 1 launch day and 1 coaching visit are needed, and that a launch duration of more than 5 days or having more than 40 coaching visits is impractical, we estimated the optimal intervention for a center with average birth volume ($z=175$) to be  a  launch duration of 2.78 days and 1 coaching visit. We also carried out optimization over all possible  combinations of  discrete values within $\mathcal{X}$, which are  $1,...,40$ coaching visits and $1,1.5,2,2.5,...,5$ for duration of intervention launch and obtained the optimal intervention as launch duration of three days with one coaching visit, $\hat{\bx}^{opt}=(3,1)$. The total cost of the estimated optimal intervention package, $\hat{\bx}^{opt}$, was \$2570.    

We calculated a 95\% confidence set for the optimal intervention $CS(\bx^{opt})$ over the grid of $\mathcal{X}$, taking all possible  numbers of coaching visits, $1,...,40$, and $1,1.5.,2,2.5,...,5$ for intervention launch duration. Out of 360 potential intervention packages, 38 (10.5\%) were included in the 95\% confidence set. The set included the following combinations: 1.5 days launch duration and 40 coaching visits; 2 days launch durations and 27 or more coaching visits; 2.5 days launch duration and less than 20 coaching visits; and 3 days launch duration and less than 5 coaching visits.   The first, second and third quartiles of the cost distribution within $CS(\bx^{opt})$ were  $Q1$=\$2462, $Q2$=\$4035, and $Q3$=\$6797.  We also  calculated 95\% simultaneous confidence bands for the probability of success under all 360 intervention compositions; plots are shown in Section 6 of the supplementary materials. For the estimated optimal intervention $\hat{\bx}^{opt}=(1,3)$, the obtained interval within the bands for the probability of oxytocin administration was  $(0.79, 0.93)$. The mean difference between the top and bottom  of the confidence band over all 360 intervention compositions was 0.07.

\section{Discussion}
 \label{Sec:Discu}
We developed the LAGO design for multiple component intervention studies with a binary outcome, where  the intervention package composition is systematically adapted as part of the design.  The  goals of studies using the LAGO design are to find the optimal intervention package, to test its effect on the  outcome of interest, and to estimate its effect as well as the effects of the individual components

The methodology in this paper  was developed for scenarios with a stagewise  analysis that does not include formal interim hypothesis testing. However, the LAGO design allows for futility stops,  since stopping the trial for futility between stages preserves the type I error. The type I error can only decrease from the nominal level when futility stops are included because when stopping for futility, the null is not rejected \citep{snapinn2006assessment}.  
 
For clear presentation of the design, methods, and theory, we focused  on a general yet practical design.  Our work opens the way for further research. For example, it would be interesting to develop methods for studies with further dependence because centers contribute data to more than one stage. The results in this paper could also be extended to continuous, count, or survival outcome data.  Finally, many design problems arise, in terms of identifying the optimal $K$, $J_k$ and $n_{jk}$ for given  settings.  
  
Many large effectiveness and implementation trials fail because current design methodology does not permit adaptation in the face of implementation failure as in, for example, the BetterBirth \citep{semrau2017outcomes} and the TasP \citep{iwuji2017universal} studies.    The LAGO design rigorously formalizes practices in public health research that are presently conducted in an ad hoc manner, with unknown consequences for the validity of the subsequent standard analysis \citep{escoffery2018scoping}. We expect widespread use of the LAGO design as a result, with potential gain for many randomized clinical trials.
   
\bibliographystyle{Chicago}
\bibliography{LearnAsYouGo}

\end{document}